\documentclass{article} % For LaTeX2e
\usepackage{iclr2024_conference,times}

\usepackage{amsmath,amsfonts,bm}

\def\eqref#1{equation~\ref{#1}}
\def\1{\bm{1}}

\DeclareMathAlphabet{\mathsfit}{\encodingdefault}{\sfdefault}{m}{sl}
\SetMathAlphabet{\mathsfit}{bold}{\encodingdefault}{\sfdefault}{bx}{n}

\usepackage{booktabs}
\usepackage{multirow}
\usepackage{multicol}
\usepackage{hyperref}
\usepackage{graphicx}
\usepackage{makecell}
\usepackage{url}
\usepackage{bbm}

\title{Solving Diffusion ODEs with \\Optimal Boundary Conditions for \\Better Image Super-Resolution}

\author{
    Yiyang Ma\textsuperscript{1}, Huan Yang\textsuperscript{2}, Wenhan Yang\textsuperscript{3}, Jianlong Fu\textsuperscript{2}, Jiaying Liu\textsuperscript{1}\thanks{Corresponding author.}\\
    \textsuperscript{1}Wangxuan Institute of Computer Technology, Peking University, 
    \textsuperscript{2}Microsoft Research,\\
    \textsuperscript{3}Pengcheng Laboratory\\
    \tt\small 
    \textsuperscript{1}\{myy12769, liujiaying\}@pku.edu.cn, \\
    \tt\small
    \textsuperscript{2}\{huayan, jianf\}@microsoft.com, \textsuperscript{3}yangwh@pcl.ac.cn
}

\newcommand{\ie}{\textit{i}.\textit{e}., }
\newcommand{\eg}{\textit{e}.\textit{g}., }
\newcommand{\wh}[1]{\textcolor{black}{#1}}

\newcommand{\myy}[1]{\textcolor{black}{#1}}
\newcommand{\chat}[1]{\textcolor{black}{#1}}
\newcommand{\red}[1]{\textcolor{red}{#1}}
\newcommand{\blue}[1]{\textcolor{blue}{#1}}

\iclrfinalcopy % Uncomment for camera-ready version, but NOT for submission.
\begin{document}

\maketitle

\begin{abstract}
Diffusion models, as a kind of powerful generative model, have given impressive results on the image super-resolution (SR) tasks. 
However, due to the randomness introduced in the reverse process of diffusion models, the performances of diffusion-based SR models fluctuate at every time of sampling, especially for the samplers with few resampled steps. 
This inherent randomness of diffusion models results in ineffectiveness and instability, making it challenging for users to guarantee the quality of SR results. 
However, our work takes this randomness as an opportunity: fully analyzing and leveraging it to the construction of an effective plug-and-play sampling method that has the potential to benefit a series of diffusion-based SR methods.
More in detail, we propose to steadily sample high-quality SR images from pre-trained diffusion-based SR models by solving diffusion ordinary differential equations (\textit{diffusion ODE}s) with optimal boundary conditions (BCs) and analyze the characteristics between the choices of BCs and their corresponding SR results. 
Our analysis shows the route to obtain an approximately optimal BC via an efficient exploration in the whole space.
The quality of SR results sampled by the proposed method with fewer steps outperforms the quality of results sampled by current methods with randomness from the same pre-trained diffusion-based SR model, which means that our sampling method ``boosts'' current diffusion-based SR models without any additional training. 
\end{abstract}

\section{Introduction}

\begin{figure}[htbp]
    \centering
    \includegraphics[width=0.9\linewidth]{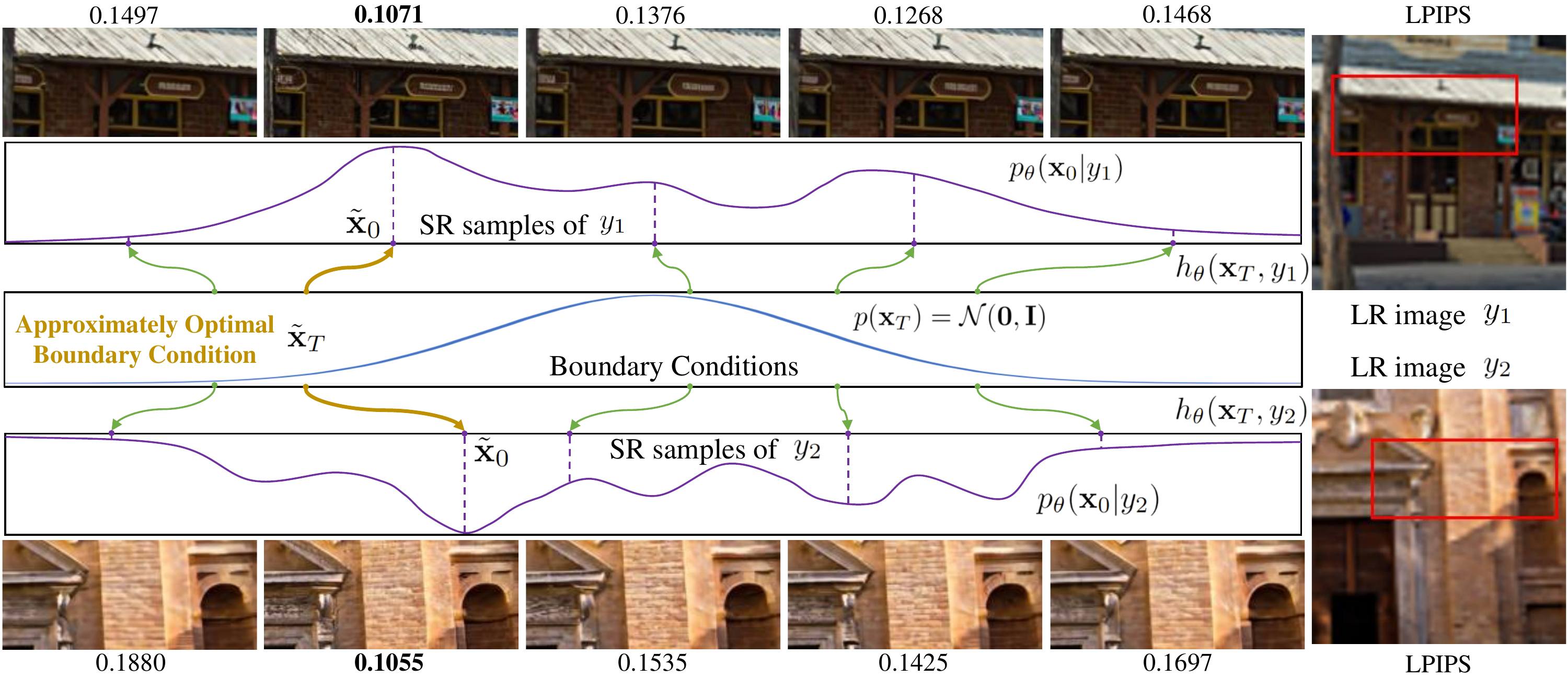}
    \caption{
    Given a well-trained diffusion-based SR model, by solving \textit{diffusion ODE}s, we can sample reasonable SR results with different BCs $\mathbf{x}_T$ as the figure shows. However, there is instability in the performances of each BC $\mathbf{x}_T$. We manage to find an approximately optimal BC $\tilde{\mathbf{x}}_T$ which can be projected to the sample $\tilde{\mathbf{x}}_0$ with nearly the highest probability density by the solution $h_\theta(\tilde{\mathbf{x}}_T, \mathbf{y})$ to \textit{diffusion ODE}. Based on our analysis in the Sec.~\ref{subsec: analyzing}, $\tilde{\mathbf{x}}_T$ is shared by different LR images $\mathbf{y}_i$. The method of finding $\tilde{\mathbf{x}}_T$ refers to the Sec.~\ref{subsec: approximating}
    \textbf{[Zoom in for best view]}}
    \label{fig: teaser}
\end{figure}

Diffusion models \citep{2020DDPM} have drawn great research attention within the domain of computer vision because of their great capacity for image generation.
Therefore, it is intuitive to leverage such powerful models to tackle the demanding task of image super-resolution (SR).
The diffusion-based image SR task is modeled as generating high-quality images by diffusion models conditioned on corresponding low-resolution images \citep{2021SR3, 2021SRDiff, 2023ResDiff, 2023SR3+}. 
However, the reverse process (\ie generating process) of diffusion models, including randomness \citep{2020DDPM, 2020DDIM, 2020ScoreModel}, leads to unstable performances of the diffusion-based SR methods.
In other words, the users cannot guarantee the quality of SR results if they lack a principled approach and can only rely on random sampling from diffusion-based models.
%
%Existing methods do not treat the randomness.
The previous methods did not consider or explore the issue of randomness.
%
%Although one can generate reasonable SR images using well-trained diffusion-based SR models by sampling methods with randomness, there is still a gap between the quality of randomly sampled results and the optimal quality. 
%
Although multiple random samplings methods can lead to reasonable SR images using well-trained diffusion-based SR models. However, we cannot guarantee the quality of one-time sampling, and the sampled results on average still fall short of optimal quality, with significant performance gaps.
%
%So, how to steadily sample SR images from pre-trained diffusion model achieving better quality makes the existing models exert themselves.
Thus, it is critical to pursue a stable sampling method that generates SR images from pre-trained diffusion models with guaranteed good performances.

Most current diffusion-based SR works \citep{2021SR3, 2021SRDiff, 2023ResDiff, 2023SR3+, 2023StableSR} focus on the model design instead of the sampling method.
The most commonly used sampling method for diffusion-based SR works is a resampled DDPM sampler with 100 steps (DDPM-100) instead of the original DDPM sampler with 1000 steps of the training noise schedule (DDPM-1000),
%because the original one is ultra time-costing despite the loss of quality of SR images. 
due to its significantly reduced time cost, despite the trade-off in SR image quality.
It is first introduced by SR3 \citep{2021SR3} from WaveGrad \citep{2020WaveGrad}. Later works following SR3 use DDPM-100 as a default setting. 
These discrete-time DDPM samplers sample from a Gaussian distribution with learned parameters at each step, resulting in instability.
\cite{2020ScoreModel} demonstrate that such discrete-time DDPM samplers can be regarded as solving diffusion stochastic differential equations (\textit{diffusion SDE}s) and further give ordinary differential equations which share the same marginal probability densities as \textit{diffusion SDE}s.
Such ordinary differential equations are referred to as \textit{diffusion ODE}s. Different from \textit{diffusion SDE}s, given a boundary condition (BC) $\mathbf{x}_T$, one can solve the \textit{diffusion ODE}s via ODE samplers (\eg DDIM \citep{2020DDIM}, DPM Solver \citep{2022DPM-Solver}) to get an exact solution $\mathbf{x}_0$.
%
%Nevertheless, the BCs $\mathbf{x}_T\sim\mathcal{N}(\mathbf{0}, \mathbf{I})$ still consist of randomness. Using random BCs $\mathbf{x}_T$ to sample SR images still suffers the problem of instability. 
%
Nevertheless, the BCs $\mathbf{x}_T\sim\mathcal{N}(\mathbf{0}, \mathbf{I})$ also come with randomness and lead to the instability issue in sampling SR images.
Hence, it is highly desirable to obtain a principled way for estimating the optimal BC $\mathbf{x}^*_T$ to steadily offer sampled SR images with high quality.

In this paper, we analyze the characteristics of the optimal BC $\mathbf{x}^*_T$ of \textit{diffusion ODE}s of SR models and propose an approach to approximate the optimal BC $\tilde{\mathbf{x}}_T$ by exploring the whole space with the criterion of a reference set containing $R$ HR-LR image pairs $\mathcal{R} = \{(\mathbf{z}_i, \mathbf{y}_i)\}_{i=1}^R$ which is a small subset of the training dataset. 
Then, we can steadily generate high-quality SR images by solving the \textit{diffusion ODE}s of the trained diffusion-based SR model with the above derived approximately optimal BC $\tilde{\mathbf{x}}_T$.
%
%We state that the optimal BC $\mathbf{x}^*_T$ used to solve the \textit{diffusion ODE} of diffusion-based SR models is independent with the LR image inputs. 
We establish that the optimal boundary condition $\mathbf{x}^*_T$ utilized to solve the \textit{diffusion ODE}s in diffusion-based SR models is independent of the LR image inputs.
Thus, we only need to prepare the approximately optimal BC $\tilde{\mathbf{x}}_T$ once to sample SR images of other unseen LR images.
%
%The experiment demonstrates that, this simple independence assumption can have already offer impressive performance empirically in plug and play manner.
The experiment demonstrates that this simple independence assumption empirically offers impressive performance in a plug-and-play manner.
The main idea is shown in Fig.~\ref{fig: teaser}.

\myy{We evaluate our method on both bicubic-SR and real-SR \wh{degradation settings}. For bicubic-SR, we train a vanilla diffusion-based SR model which simply concatenates LR images with noisy images $\mathbf{x}_t$ as the architecture proposed in SR3 \citep{2021SR3}. For real-SR, we apply our method to StableSR \citep{2023StableSR}, which finetunes pre-trained Stable Diffusion \citep{2022StableDiffusion} on real-SR data.}
% In order to evaluate the effectiveness of our method, we train a vanilla diffusion-based SR model with a noise-prediction UNet which simply concatenate LR images with noisy images $\mathbf{x}_t$ as the architecture proposed in SR3 \cite{2021SR3}. 
Experiments show that the quality of SR images sampled by few-step \textit{diffusion ODE} samplers with our explored BC $\tilde{\mathbf{x}}_T$ significantly outperforms the quality of results sampled by existing methods owning the same architecture.
%
%Our method does not have any restrict on the architecture of diffusion-based SR models. So, any diffusion-based SR models can use the proposed method to steadily sample high-quality SR images with few steps, reaching better performance. To this extent, the proposed method boosts current diffusion-based SR models.
%
Our method is not restricted to any specific architecture of diffusion-based SR models. \myy{As the models we utilize in bicubic-SR and real-SR are quite different, the \wh{versatility} of our method can be demonstrated by experiments.} Therefore, any diffusion-based SR model can leverage the proposed method to steadily sample high-quality SR images with only a few steps, and achieve improved performance. In this way, our method can ``boost'' existing diffusion-based SR models in the plug-and-play manner.

\section{Related Work}

\subsection{Image Super-Resolution}

Image super-resolution has drawn great research interest in recent years \citep{2014SRCNN, 2016VDSR, 2017DenseSR, 2017EDSR, 2017SRRestNet&SRGAN, 2018ESRGAN, 2018RCAN, 2021SWINIR}. As a pioneer work of deep-learning based SR methods, SRCNN \citep{2014SRCNN} builds a 3-layer convolutional neural network to map LR patches to SR patches with the criterion of MSE between SR patches and HR patches, getting better PSNR than traditional methods. SRResNet \citep{2017SRRestNet&SRGAN} introduces residual connections into SR networks, achieving impressive performances. RCAN \citep{2018RCAN} uses channel-attention mechanism to learn local-correlation which is crucial to the SR task. SWINIR \citep{2021SWINIR} leverages vision transformers \citep{2020ViT, 2021SWIN} to build backbones of SR neural networks and outperforms CNN-based NNs.

However, PSNR between SR images and HR images has a gap with the visual quality of SR images. Using generative models can synthesize more perceptually pleasant results. Thus, SRGAN \citep{2017SRRestNet&SRGAN} introduces GANs \citep{2014GAN} to SR tasks. Furthermore, \cite{2020PULSE, 2021GPEN, 2021GLEAN} incorporate pre-trained GANs from specific domains into their SR frameworks, leveraging the generative capabilities of these GANs. PixelSR \citep{2017PixelSR} uses auto-regressive models to generate SR images pixel-by-pixel. SRFlow \citep{2020SRFlow} models SR tasks by normalizing flow-based models \citep{2018GLOW}. SR3 \citep{2021SR3} first uses diffusion models \citep{2020DDPM, 2020ScoreModel} to generate SR images conditioned on corresponding LR images. DDRM \citep{2022DDRM} designs a training-free algorithm to guide pre-trained diffusion models to generate high-quality images which are consistent with the LR images. StableSR leverages pre-trained Stable Diffusion \citep{2022StableDiffusion} as a generative prior.

\subsection{Diffusion Models}

\chat{In recent years, diffusion models \citep{2020DDPM, 2020ScoreModel}, a type of generative model, have achieved impressive results across various \wh{research} domains}, including image generation \citep{2021ADM, 2021ImprovedDDPM}, text-to-image generation \citep{2021GLIDE, 2022DALLE2, 2022Imagen}, multi-modal generation \citep{2022MMDiffusion, 2023UMMDiffusion} and so on. Diffusion models are first proposed by \cite{2015OriginalDiffusion} and simplified as DDPM by \cite{2020DDPM} which can be trained as several simple denoising models. ImprovedDDPM \citep{2021ImprovedDDPM} proposes to learn the variance of each reverse step and AnalyticDPM \citep{2022AnalyticDPM} claims that such variances have analytic forms which not need to be learned. \cite{2020ScoreModel} extend the diffusion models with discrete Markovian chains to continuous differential equations. \cite{2022ImagenVideo} propose to train diffusion models by ``velocity'', getting more efficiency. \cite{2022StableDiffusion} build diffusion models on latent spaces instead of image spaces, reducing the training and inferring cost.

In terms of applying diffusion models, GLIDE \citep{2021GLIDE} first proposes to build a diffusion model to generate images from descriptive texts. DALL$\cdot$E 2 \citep{2022DALLE2} and Imagen \citep{2022Imagen} design better architecture and use more computing resources, achieving better performances. Palette \citep{2022Palette} first applies diffusion models to image-to-image translation tasks. DreamBooth \citep{2022DreamBooth} finetunes pre-trained text-to-image diffusion models to achieve the goal of subject-driven generation. MM-Diffusion \citep{2022MMDiffusion} generates aligned audios and videos at the same time. \cite{2022MakeaVideo} create novel videos from texts without text-to-video data. These works prove that diffusion models have strong generative abilities.

\section{Sampling SR Images with Optimal BCs of \textit{Diffusion ODE}s}
\label{sec: approach}

We first review diffusion models and their continuous differential equations, then analyze the optimal BCs $\mathbf{x}^*_T$ used by \textit{diffusion ODE}s to sample SR images from diffusion-based SR models, last depict the method of approximating the optimal BCs $\tilde{\mathbf{x}}_T$ in Eqn.~\ref{eqn: target} with the criterion of a reference set containing $R$ image pairs. With the approximately optimal $\tilde{\mathbf{x}}_T$, we can sample high-quality SR images from diffusion-based SR models by solving \textit{diffusion ODE}s steadily.

\subsection{Diffusion Models, \textit{Diffusion SDE}s and \textit{Diffusion ODE}s}

Diffusion models \citep{2020DDPM, 2020ScoreModel}, a kind of generative model, first map samples from an unknown distribution (\eg the natural image distribution) to samples from a well-known distribution (\eg the standard Gaussian distribution) by gradually adding noise, and then attempt to revert such process via denoising step by step. The first process is called \textit{forward process}. Taking $\mathbf{x}_0$ as a sample of the unknown distribution $X$, $T$ as the number of noise-adding step, the state $\mathbf{x}_t, t \in [0, T]$ of \textit{forward process} satisfies
\begin{equation}
    q(\mathbf{x}_t|\mathbf{x}_0) = \mathcal{N}(\mathbf{x}_t;\alpha(t)\mathbf{x}_0, \sigma^2(t)\mathbf{I}), q(\mathbf{x}_T) =  \mathcal{N}(\mathbf{x}_T;\mathbf{0}, \mathbf{I}),
\end{equation}
where $\alpha(t), \sigma(t)$ are differentiable functions of $t$ defined by hyper-parameters.
Furthermore, \cite{2021VariationalDPM} prove that the transition distribution $q(\mathbf{x}_t|\mathbf{x}_0)$ can be given by the following stochastic differential equation (SDE) at any $t \in [0, T]$:
\begin{equation}
    {\rm d}\mathbf{x}_t = f(t)\mathbf{x}_t{\rm d}t + g(t){\rm d}\mathbf{w}_t,
\end{equation}
where $\mathbf{w}_t$ is a standard Wiener process, and $f(t), g(t)$ are given by
\begin{equation}
    f(t) = \frac{{\rm d}\log \alpha(t)}{{\rm d}t},
    g^2(t) = \frac{{\rm d}\sigma^2(t)}{{\rm d}t} - 2\frac{{\rm d}\log \alpha(t)}{{\rm d}t}\sigma^2(t).
\end{equation}
The \textit{reverse process} attempts to learn a parameterized distribution $p_\theta(\mathbf{x}_0)$ to fit the real data distribution $q(\mathbf{x}_0)$ by using a trained noise-prediction model $\bm{\epsilon}_\theta(\mathbf{x}_t, t)$ to gradually generate $\mathbf{x}_0$ from $\mathbf{x}_T$ \citep{2020DDPM}. \cite{2022DPM-Solver} prove that the reverse process can be done by solving the following parameterized SDE (\textit{diffusion SDE}) with numerical solvers:
\begin{equation}
    {\rm d}\mathbf{x}_t = [f(t) \mathbf{x}_t + \frac{g^2(t)}{\sigma(t)}\bm{\epsilon}_\theta(\mathbf{x}_t, t)]{\rm d}t + g(t){\rm d}\bar{\mathbf{w}}_t, \mathbf{x}_T \sim \mathcal{N}(\mathbf{0}, \mathbf{I}),
\end{equation}
where $\bm{\epsilon}_\theta(\mathbf{x}_t, t)$ is a trainable noise-prediction neural network and $\bar{\mathbf{w}}_t$ is another standard Wiener process in the reverse time. The original DDPM \citep{2020DDPM} sampler used by current diffusion-based SR models is a discrete-time solver of \textit{diffusion SDE}. When discretizing \textit{diffusion SDE}s, the step sizes are limited because the Wiener process $\bar{\mathbf{w}}_t$ contains randomness. Thus, the resampled DDPM-100 sampler which is mentioned before with larger step sizes performs not satisfying.

Moreover, \cite{2020ScoreModel} give an ordinary differential equation (ODE) which has the same marginal distribution of \textit{diffusion SDE}:
\begin{equation}
\label{eqn: diffusion ODE}
    \frac{{\rm d}\mathbf{x}_t}{{\rm d}t} = f(t) \mathbf{x}_t + \frac{g^2(t)}{2\sigma(t)}\bm{\epsilon}_\theta(\mathbf{x}_t, t), \mathbf{x}_T \sim \mathcal{N}(\mathbf{0}, \mathbf{I}).
\end{equation}
Such ODE is called \textit{diffusion ODE}.
Because \textit{diffusion ODE}s have no randomness, one can get an exact solution $\mathbf{x}_0$ given a BC $\mathbf{x}_T$ by solving the \textit{diffusion ODE}s with corresponding numerical solvers like DDIM \citep{2020DDIM} or DPM-Solver \citep{2022DPM-Solver}. Thus, we can use a parameterized projection: 
\begin{equation}
\label{eqn: diffusion ODE projection}
    \mathbf{x}_0 = h_\theta(\mathbf{x}_T), \mathbf{x}_T \sim \mathcal{N}(\mathbf{0}, \mathbf{I}),
\end{equation}
to represent the solution of \ref{eqn: diffusion ODE}. We can extend the diffusion models to conditional ones $p_\theta(\mathbf{x}_0|c)$ by providing conditions $c$ when training the noise-prediction model $\bm{\epsilon}_\theta(\mathbf{x}_t, c, t)$. By randomly dropping the conditions \myy{during the training process}, the model can be jointly conditional and unconditional \citep{2021ClassiferFreeGuidance}. We define the projections:
\begin{equation}
\label{eqn: conditional diffusion ODE projection}
    \mathbf{x}_0 = h_\theta(\mathbf{x}_T, c), \mathbf{x}_0 = h_\theta(\mathbf{x}_T, \phi),
\end{equation} 
are the solution to conditional \textit{diffusion ODE} and the solution to unconditional \textit{diffusion ODE} of the same diffusion model respectively, where $\phi$ denotes the blank condition which is dropped.
% \begin{equation}
% \label{eqn: unconditional diffusion ODE projection}
%     \mathbf{x}_0 = h_\theta(\mathbf{x}_T, \phi),
% \end{equation} 
% is the solution to unconditional \textit{diffusion ODE} of the same diffusion model.

\subsection{Analyzing Optimal BCs \texorpdfstring{$\mathbf{x}^*_T$}{} of \textit{Diffusion ODE}s for Diffusion-based SR Models}
\label{subsec: analyzing}
For image SR tasks, steady SR results mean deterministic samples of the learned conditional distribution $p_\theta(\mathbf{x}_0|c)$, where the conditions $c$ are LR images $\mathbf{y}$. In other words, we should only sample once from the distribution. The parameterized distribution $p_\theta(\mathbf{x}_0|\mathbf{y})$ learned by a well-trained diffusion model is a fitting to the data probability distribution $q(\mathbf{x}_0|\mathbf{y})$ and the training data pairs $(\mathbf{z}_i,\mathbf{y}_i)$ are samples and conditions of the distribution $q(\mathbf{x}_0|\mathbf{y})$, where $\mathbf{z}_i$ denotes the corresponding HR image of $\mathbf{y}_i$. From the perspective of max-likelihood, the $(\mathbf{z}_i, \mathbf{y}_i)$ pairs should be located at the point with the biggest probability distribution $q(\mathbf{x}_0| y_1)$:
\begin{equation}
\label{eqn: HR-LR max likelihood}
    \mathbf{z}_i = \mathop{\arg\max}_{\mathbf{x}_0}{q(\mathbf{x}_0|\mathbf{y}_i)}.
\end{equation}
So, the optimal sample of $p_\theta(\mathbf{x}_0|\mathbf{y})$ should satisfy:
\begin{equation}
\label{eqn: optimal x_0}
    \mathbf{x}^*_0 = \mathop{\arg\max}_{\mathbf{x}_0}{p_\theta(\mathbf{x}_0|\mathbf{y})}.
\end{equation}
When we solve \textit{diffusion ODE}s to sample from the diffusion model $p_\theta(\mathbf{x}_0|\mathbf{y})$, we actually sample $\mathbf{x}_T \sim \mathcal{N}(\mathbf{0}, \mathbf{I})$ and project $\mathbf{x}_T$ to final samples $\mathbf{x}_0$ via the projection in Eqn.~\ref{eqn: conditional diffusion ODE projection}. By leveraging the law of total probability, we can replace the variable from $\mathbf{x}_0$ to $\mathbf{x}_T$, getting the likelihood of $\mathbf{x}_T$:
% by replacing the variable $\mathbf{x}_0$ with $\mathbf{x}_T$, we have:
\begin{equation}
\label{eqn: re-sample}
    p_\theta'(\mathbf{x}_T|\mathbf{y}) =  \sum_{\bar{\mathbf{y}}\in\mathcal{C}} {p_\theta(\mathbf{x}_0|\mathbf{y})|_{\mathbf{x}_0 = h_\theta(\mathbf{x}_T, \bar{\mathbf{y}})}p(\bar{\mathbf{y}})}
    = p_\theta(h_\theta(\mathbf{x}_T, \mathbf{y})),
\end{equation}
where $\mathcal{C}$ is the theoretically universal set of all LR images and $\bar{\mathbf{y}}$ indicates all the LR images. The proof of Eqn.~\ref{eqn: re-sample} refers to the Sec.~\ref{suppsec: proof} of the appendix.
For unconditional sampling, we have:
\begin{equation}
\label{eqn: re-sample unconditional}
    p_\theta'(\mathbf{x}_T) = p_\theta(\mathbf{x}_0)|_{\mathbf{x}_0 = h_\theta(\mathbf{x}_T, \phi)} = p_\theta(h_\theta(\mathbf{x}_T, \phi)).
\end{equation}
% It should be noticed that $p_\theta(\mathbf{x}_0|\mathbf{y})$ is a probability distribution of $\mathbf{x}_0$ while $p_\theta(h_\theta(\mathbf{x}_T, \mathbf{y}))$ is a probability distribution of $\mathbf{x}_T$. For simplicity, we do not change the symbols. 
By substituting Eqn.~\ref{eqn: re-sample} into Eqn.~\ref{eqn: optimal x_0}, optimal BCs and samples should satisfy:
\begin{equation}
\label{eqn: optimal x_T}
    \mathbf{x}^*_T = \mathop{\arg\max}_{\mathbf{x}_T \sim \mathcal{N}(\mathbf{0}, \mathbf{I})}{p_\theta(h_\theta(\mathbf{x}_T, \mathbf{y}))}, \mathbf{x}^*_0 = h_\theta(\mathbf{x}^*_T, \mathbf{y}).
\end{equation}
% Because a well-trained $p_\theta(\mathbf{x}_0|\mathbf{y})$ is a fitting to $q(\mathbf{x}_0|\mathbf{y})$. 
Based on Bayesian rule, we have:
\begin{equation}
\label{eqn: bayesian rule}
    p_\theta(\mathbf{x}_0|\mathbf{y})
    =
    \frac{p_\theta(\mathbf{x}_0, \mathbf{y})}{p(\mathbf{y})}
    =
    \frac{p_\theta(\mathbf{y}|\mathbf{x}_0)}{p(\mathbf{y})}p_\theta(\mathbf{x}_0).
\end{equation}
By replacing the variable from $\mathbf{x}_0$ to $\mathbf{x}_T$ in Eqn.~\ref{eqn: bayesian rule} with Eqn.~\ref{eqn: re-sample} and Eqn.~\ref{eqn: re-sample unconditional}, the parameterized conditional distribution is:
\begin{equation}
\label{eqn: post x_0}
    p_\theta(h_\theta(\mathbf{x}_T, \mathbf{y})) = \frac{p_\theta(\mathbf{y}|h_\theta(\mathbf{x}_T, \phi))}{p(\mathbf{y})}p_\theta(h_\theta(\mathbf{x}_T, \phi)).
\end{equation}
In Eqn.~\ref{eqn: post x_0}, $p(\mathbf{y})$ is the prior probability distribution of LR images which is a uniform distribution, and $p_\theta(h_\theta(\mathbf{x}_T, \phi))$ is not related to the LR image $\mathbf{y}$. $p_\theta(\mathbf{y}|h_\theta(\mathbf{x}_T, \phi))$ is an implicit classifier, \myy{indicating the probability of image $\mathbf{y}$ is the corresponding LR image of an unconditionally generated image $h_\theta(\mathbf{x}_T, \phi)$. For a well-trained model, such probability is also approximately uniform, because the distribution of unconditionally generated images will be approximate to the real distribution of choosing images in the dataset, which is uniform.} Thus, $p_\theta(h_\theta(\mathbf{x}_T, \mathbf{y}))$ is approximately independent to the specific LR images $\mathbf{y}$, which indicates:
\begin{equation}
\label{eqn: optimal x_T conclusion}
    \mathbf{x}^*_T 
    =
    \mathop{\arg\max}_{\mathbf{x}_T \sim \mathcal{N}(\mathbf{0}, \mathbf{I})}{p_\theta(h_\theta(\mathbf{x}_T, \mathbf{y}))} 
    \approx
    \mathop{\arg\max}_{\mathbf{x}_T \sim \mathcal{N}(\mathbf{0}, \mathbf{I})}{p_\theta(h_\theta(\mathbf{x}_T, \mathbf{y}_i))}, \forall \mathbf{y}_i \in \mathcal{C}.
\end{equation}
We design an experiment in the Sec.~\ref{suppsec: validation} of the appendix to validate a derivation of the approximate independence of $p_\theta(h_\theta(\mathbf{x}_T, \mathbf{y}))$ to different $\mathbf{y}$. Hitherto, we have stated that the optimal BC $\mathbf{x}^*_T$ is approximately general for different LR images $\mathbf{y}$. In the next subsection, we depict how to approximate $\mathbf{x}^*_T$ with the criterion of a reference set containing $R$ HR-LR image pairs $\mathcal{R} = \{(\mathbf{z}_i, \mathbf{y}_i)\}_{i=1}^R$ which is a subset of the training dataset.

\subsection{Approximating Optimal BCs \texorpdfstring{$\tilde{\mathbf{x}}_T$}{} of \textit{Diffusion ODE}s for Diffusion-based SR Models}
\label{subsec: approximating}

As we have discussed before, a well-trained model $p_\theta(\mathbf{x}_0|\mathbf{y})$ is a fitting of $q(\mathbf{x}_0|\mathbf{y})$. Thus, we can take $q(\mathbf{x}_0|\mathbf{y})$ to substitute $p_\theta(\mathbf{x}_0|\mathbf{y})$ in Eqn.~\ref{eqn: optimal x_T conclusion}, getting an approximation $\tilde{\mathbf{x}}_T$ of $\mathbf{x}^*_T$:
\begin{equation}
    \tilde{\mathbf{x}}_T
    =
    \mathop{\arg\max}_{\mathbf{x}_T \sim \mathcal{N}(\mathbf{0}, \mathbf{I})}{q(h_\theta(\mathbf{x}_T, \mathbf{y}_i))}.
\end{equation}
Besides, we have the max-likelihood Eqn.~\ref{eqn: HR-LR max likelihood} of $q(\mathbf{x}_0|\mathbf{y})$:
\begin{equation}
\label{eqn: max}
    \mathbf{z}_i = \mathop{\arg\max}_{\mathbf{x}_0}{q(\mathbf{x}_0|\mathbf{y}_i)} = h_\theta(\mathop{\arg\max}_{\mathbf{x}_T \sim \mathcal{N}(\mathbf{0}, \mathbf{I})}{q(h_\theta(\mathbf{x}_T, \mathbf{y}_i))}, \mathbf{y}_i).
\end{equation}
Considering the characteristics of natural images, the distribution $q(\mathbf{x}_0|\mathbf{y})$ is a continuous distribution. So, there exists a neighbour around $\mathbf{z}_i$ where $q(\mathbf{x}_0|\mathbf{y}_i)$ is monotonic. Furthermore, the closer $\mathbf{x}_0$ gets to $\mathbf{z}_i$, the bigger $q(\mathbf{x}_0|\mathbf{y}_i)$ is. By taking $M(\cdot, \cdot)$ as the function which measures the distance of two images, the $\tilde{\mathbf{x}}_T$ can be approximated by:
\begin{equation}
    \tilde{\mathbf{x}}_T
    =
    \mathop{\arg\max}_{\mathbf{x}_T \sim \mathcal{N}(\mathbf{0}, \mathbf{I})}{q(h_\theta(\mathbf{x}_T, \mathbf{y}_i))}
    \approx
    \myy{\mathop{\arg\min}_{\mathbf{x}_T \sim \mathcal{N}(\mathbf{0}, \mathbf{I})}{M(h_\theta(\mathbf{x}_T, \mathbf{y}_i), \mathbf{z}_i)}
    .}
\end{equation}
Because the monotonicity of $q(\mathbf{x}_0|\mathbf{y}_i)$ is limited in a small neighbour, we can use a set containing $R$ HR-LR image pairs $\mathcal{R} = \{(\mathbf{z}_i, \mathbf{y}_i)\}_{i=1}^R$ to calculate $\tilde{\mathbf{x}}_T$ \myy{to achieve better approximation}:
\begin{equation}
\label{eqn: analitical target}
    \tilde{\mathbf{x}}_T
    \approx
    \mathop{\arg\min}_{\mathbf{x}_T \sim \mathcal{N}(\mathbf{0}, \mathbf{I})}{\sum_{i = 1}^{R}{M(h_\theta(\mathbf{x}_T, \mathbf{y}_i), \mathbf{z}_i)}}.
\end{equation}
Considering the perceptual characteristics of images, we take LPIPS \citep{2018LPIPS} as the implementation of $M(\cdot, \cdot)$. Because the projection $h_\theta$ is the solution to \textit{diffusion ODE}, it is difficult to give an analytical result of Eqn.~\ref{eqn: analitical target}. We use the idea of the Monte Carol method to estimate $\tilde{\mathbf{x}}_T$. We randomly sample $K$ $\mathbf{x}_T \sim \mathcal{N}(\mathbf{0}, \mathbf{I})$, calculate Eqn.~\ref{eqn: analitical target} and choose the best one:
\begin{equation}
\label{eqn: target}
    \tilde{\mathbf{x}}_T
    \approx
    \mathop{\arg\min}_{\mathbf{x}_T \in \mathcal{K}}{\sum_{i = 1}^{R}{{\rm LPIPS}(h_\theta(\mathbf{x}_T, \mathbf{y}_i), \mathbf{z}_i)}},
\end{equation}
where $\mathcal{K}$ is the set of randomly sampled $K$ $\mathbf{x}_T \sim \mathcal{N}(\mathbf{0}, \mathbf{I})$. Last, given unseen LR images $\mathbf{y}$, the corresponding SR images can be generated by:
\begin{equation}
    \tilde{\mathbf{x}}_0 = h_\theta(\tilde{\mathbf{x}}_T, \mathbf{y}).
\end{equation}
\section{Experiments}

In order to demonstrate the effectiveness of the proposed sampling method, we \myy{apply our method on two diffusion-based SR models. For bicubic-SR, we train a vanilla model following SR3 \citep{2021SR3} as the baseline. For real-SR, we utilize StableSR \citep{2023StableSR} with $w=0.5$ without any color fixing and DiffIR \citep{2023DiffIR} as the baseline. It is noted that DiffIR officially employs a 4-step-DDPM sampler without noise which we call ``D-4'' as shown in Tab.~\ref{tab: main real qualitative results}.}
%train a vanilla diffusion-based SR model which has similar architecture of SR3 \cite{2021SR3} as a baseline and evaluate several commonly-used sampling methods and our method on it.

\subsection{Implementation Details}

\textbf{Datasets.} \myy{For bicubic-SR,} we train the model on the widely-used dataset DF2k~\citep{2017DIV2K, 2017EDSR} which containing 3,450 high-resolution images. We train a 64$\times$64 $\rightarrow$ 256$\times$256 model. The training and architecture details of the bicubic-SR model refer to the Sec.~\ref{suppsec: implementation} of the appendix. \myy{For real-SR, we directly leverage the official pre-trained model of StableSR \citep{2023StableSR} with $w=0.5$ without any color fixing.}

\myy{To test the performances of bicubic-SR,} we use 3 different datasets containing DIV2k-test~\citep{2017DIV2K}, Urban100 \citep{2015Urban100}, B100 \citep{2001B100}.
For DIV2k-test and Urban100, we randomly crop 1,000 256$\times$256 patches as HR images and downscale them to 64$\times$64 patches by bicubic kernel as corresponding LR patches. 
%
%For B100, we randomly crop 200 patches because the image resolutions are not big in this dataset.
%
For B100, we randomly extract 200 patches as the image resolutions in this dataset are not large compared with those in other datasets.

\myy{To test the performances of real-SR, we utilize datasets including DIV2k-test \citep{2017DIV2K} and RealSR \citep{2019RealSR_Set}. For DIV2k-test, we employ the $\times4$ degradation process proposed by \cite{2021RealESRGAN}, synthesizing 1,000 128$\times$128 LR patches. For RealSR, we randomly crop 1,000 128$\times$128 $\rightarrow$ 512$\times$512 LR-HR pairs.}

% \textbf{Training details.} 
% %
% Following SR3~\cite{2021SR3}, we build a UNet-based noise-predition model which directly concatenates LR images with noisy states $\mathbf{x}_t$ along the channel dimension for our diffusion model and upsample origin LR images to the size of SR images by bicubic kernel to ensure that they have the same sizes as $\mathbf{x}_t$.
% %
% Our UNet has similar architecture to the one used by SR3, but only contains about 36M parameters. We train the model for 2M iterations with a batch size of 16 at first, then train the model for another 1M iterations with a batch size of 64. The learning rate is fixed to $1e-4$.
% %
% More details of the UNet and the diffusion model can be found in supplementary details.

\textbf{Compared methods and metrics.} 
This paper proposes a method of sampling from diffusion-based SR models, so, the main baselines are current sampling methods used by other diffusion-based SR models on the same model. \myy{For bicubic-SR,} we leverage several resampled DDPM samplers and \textit{diffusion ODE} samplers.
It is noted that we report the performances of DDPM-1000 \citep{2020DDPM} as upper bounds of previous sampling methods, which serves as evidence of our model's capability.
\myy{For real-SR, we employ resampled DDPM-200 following the official setting of StableSR \citep{2023StableSR} and DDIM-50 \citep{2020DDIM} as the baseline of \textit{diffusion ODE} solver. We utilize PSNR on RGB channels and LPIPS \citep{2018LPIPS} as evaluation metrics. For real-SR, we further adopt DISTS \citep{2020DISTS} to demonstrate the generality of the proposed method on diverse perceptual metrics.}
\begin{table}[tbp]
\vspace{-15mm}
    \caption{
    Qualitative results \myy{of bicubic-SR} on test datasets. ``$\tilde{\mathbf{x}}_T$'' denotes ``approximately optimal boundary condition'' calculated by the proposed method. 
    The metrics of the bottom 9 rows are all sampled with \underline{the same SR3 model} trained by us. \myy{``DPMS'' denotes DPM-Solver \citep{2022DPM-Solver}.} Numbers of PSNR are calculated on RGB channels.
    \red{Red} numbers denote the best performances and \blue{blue} numbers denote the second best performances.}
    \centering
    \begin{tabular}{c|c|cc|cc|cc}
        \toprule
         \multicolumn{2}{c|}{\multirow{2}{*}{Model (\& sampling method)}} & \multicolumn{2}{c|}{DIV2k-test} & \multicolumn{2}{c|}{Urban100} & \multicolumn{2}{c}{BSD100}
         \\
         \multicolumn{2}{c|}{} & LPIPS~$\downarrow$ & PSNR~$\uparrow$ & LPIPS~$\downarrow$ & PSNR~$\uparrow$ & LPIPS~$\downarrow$ & PSNR~$\uparrow$
         \\
         \midrule
         % \multicolumn{2}{c|}{Bicubic} & 0.4065 & 28.50 & 0.4826 & 21.75 & 0.5282 & \red{24.18} 
         % \\
         % \midrule
         \multicolumn{2}{c|}{ESRGAN} & 0.1082 & 28.18 & 0.1226 & 23.04 & 0.1579 & 23.65
         \\
         \multicolumn{2}{c|}{RankSRGAN} & 0.1171 & 27.98 & 0.1403 & 23.16 & 0.1714 & 23.80
         \\
         \multicolumn{2}{c|}{SRDiff} & 0.1286 & \textcolor{blue}{28.96} & 0.1391 & 23.88 & 0.2046 & \red{24.17}
         \\
         \midrule
         \multirow{10}{*}{SR3}& DDPM-1000 & 0.1075 & 28.75 & 0.1165 & 24.33 & 0.1555 & 23.86
         \\
         & DDPM-250 & 0.1142 & 28.95 & 0.1181 & \textcolor{blue}{24.41} & 0.1621 & 24.00
         \\
         & DDPM-100 & 0.1257 & \textcolor{red}{29.16} & 0.1232 & \textcolor{red}{24.51} & 0.1703 & \blue{24.15}
         \\
         \cmidrule(r){2-8}
         & DPMS-20 & 0.1653 & 27.25 & 0.1413 & 23.46 & 0.2037 & 22.79
         \\
         & DDIM-50 & 0.1483 & 28.55 & 0.1333 & 24.16 & 0.1823 & 23.75
         \\
         & DDIM-100 & 0.1571 & 28.16 & 0.1335 & 24.05 & 0.1950 & 23.55
         \\
         \cmidrule(r){2-8}
         & DPMS-20 + $\tilde{\mathbf{x}}_T$ & 0.1210 & 27.45 & 0.1179 & 23.57 & 0.1687 & 22.81
         \\
         & DDIM-50 + $\tilde{\mathbf{x}}_T$ & \blue{0.1053} & 28.65 & \blue{0.1164} & 24.26 & \blue{0.1552} & 23.99
         \\
         & DDIM-100 + $\tilde{\mathbf{x}}_T$ & \red{0.1032} & 28.48 & \red{0.1136} & 24.12 & \red{0.1505} & 23.67
         \\
         \bottomrule
    \end{tabular}
    \label{tab: main bicubic qualitative results}
\end{table}

\begin{table}[tbp]
\vspace{-5mm}
    \caption{
    Qualitative results of \myy{real-SR} on test datasets. ``$\tilde{\mathbf{x}}_T$'' denotes ``approximately optimal boundary condition'' calculated by the proposed method. 
    \myy{The metrics of the bottom 3 rows are all sampled with \underline{the same StableSR model} \citep{2023StableSR}.
    Numbers of PSNR are calculated on RGB channels.}
    \textcolor{red}{Red} numbers denote the best performances and \textcolor{blue}{blue} numbers denote the second best performances.}
    \centering
    \begin{tabular}{c|c|ccc|ccc}
        \toprule
         \multicolumn{2}{c|}{\multirow{2}{*}{Model (\& sampling method)}} & \multicolumn{3}{c|}{DIV2k-test} & \multicolumn{3}{c}{RealSR}
         \\
         \multicolumn{2}{c|}{} & DISTS~$\downarrow$ & LPIPS~$\downarrow$ & PSNR~$\uparrow$ & DISTS~$\downarrow$ & LPIPS~$\downarrow$ & PSNR~$\uparrow$
         \\
         \midrule
         % \multicolumn{2}{c|}{Bicubic} & 0.3293 & 0.7047 & \red{23.08} & 0.2659 & 0.4477 & \red{26.42}
         % \\
         % \midrule
         \multicolumn{2}{c|}{RealSR} & 0.3051 & 0.5148 & 22.52 & 0.2532 & 0.3673 & \blue{26.30}
         \\
         \multicolumn{2}{c|}{BSRGAN} & 0.2253 & 0.3416 & 22.13 & 0.2057 & 0.2582 & 25.52
         \\
         \multicolumn{2}{c|}{DASR} & 0.2340 & 0.3444 & 22.02 & 0.2113 & 0.3014 & \red{26.32}
         \\
         \multicolumn{2}{c|}{Real-ESRGAN} & 0.2108 & 0.3109 & 22.36 & 0.2020 & 0.2511 & 25.12
         \\
         \multicolumn{2}{c|}{KDSR-GAN} & 0.2022 & 0.2840 & 22.92 & \blue{0.2006} & \blue{0.2425} & 26.09
         \\
         \midrule
         \multirow{3}{*}{StableSR}& DDPM-200 & 0.2010 & 0.3189 & 19.42 & 0.2210 & 0.3065 & 21.37
         \\
         % \cmidrule(r){2-8}
         & DDIM-50 & 0.2217 & 0.3629 & 18.82 & 0.2336 & 0.3536 & 21.24
         \\
         % \cmidrule(r){2-8}
         & DDIM-50 + $\tilde{\mathbf{x}}_T$ & 0.2046 & 0.3169 & 19.55 & 0.2164 & 0.2999 & 22.13
         \\
         \midrule
         \multirow{2}{*}{DiffIR}& D-4 & \blue{0.1773} & \blue{0.2360} & \blue{22.94} & 0.2076 & 0.2604 & 25.33
         \\
         % \cmidrule(r){2-8}
         & D-4 + $\tilde{\mathbf{x}}_T$ & \red{0.1772} & \red{0.2357} & \red{22.95} & \red{0.1993} & \red{0.2419} & 25.82
         \\
         \bottomrule
    \end{tabular}
    \label{tab: main real qualitative results}
\vspace{-5mm}
\end{table}
Besides, we report the performances of other SOTA SR methods. \myy{For bicubic-SR,} we show the performances of SRDiff \citep{2021SRDiff}, and GAN-based methods including ESRGAN \citep{2018ESRGAN} and RankSRGAN \citep{2019RankSRGAN}. \myy{For real-SR, we show the performances of RealSR \citep{2020RealSR}, BSRGAN \citep{2021BSRGAN}, Real-ESRGAN \citep{2021RealESRGAN}, DASR \citep{2022DASR}, and KDSR-GAN \citep{2023KDSR}. We use the open-source codes and pre-trained models of these methods without any modification. The details of the source codes of the compared methods refer to the Sec.~\ref{suppsec: code source} of the appendix.}
%
%To the best of our knowledge, we are the first to outperform such GAN-based \cite{2014GAN} SR models by diffusion-based models even if the number of the our model's parameter is small.
To the best of our knowledge, the diffusion-based models with $\tilde{\mathbf{x}}_T$ have achieved superior performances compared to GAN-based SR models. % even with smaller numbers of parameters. 
This highlights the effectiveness and efficiency of our method in surpassing the capabilities of GAN-based SR models.

\textbf{Settings of calculating $\tilde{\mathbf{x}}_T$ and \textit{diffusion ODE} solvers.} 
As we have discussed in the Sec.~\ref{subsec: approximating}, we use a reference set $\mathcal{R} = \{(\mathbf{z}_i, \mathbf{y}_i)\}_{i=1}^R$ which contains HR-LR image pairs and a set $\mathcal{K}$ which contains $K$ randomly sampled $\mathbf{x}_T$ to calculate the approximately optimal BC $\tilde{\mathbf{x}}_T$. In practice, the $R$ and $K$ are set to 300 and 1,000 respectively for both bicubic-SR and real-SR. The reference sets are synthesized from DIV-2k training set by utilizing the degradation used in the training process.
%
% In practice, we randomly crop $R=300$ 256$\times$256 patches from DF2K dataset as reference HR patches, downsample them to 64$\times$64 as reference LR patches and set $K=1000$. 
%
The discussion on the effect of $R$ and $K$ refers to the Sec.~\ref{subsec: ablation}. 
For \textit{diffusion ODE} solvers, we use DDIM \citep{2020DDIM} \myy{on real-SR and further adopt DPM-Solver \citep{2022DPM-Solver} on bicubic-SR to demonstrate that our method can be generally employed to different types of \textit{diffusion ODE} solvers.}

\begin{figure}[tp]
    \centering
    \vspace{-5mm}
    \includegraphics[width=0.95\linewidth]{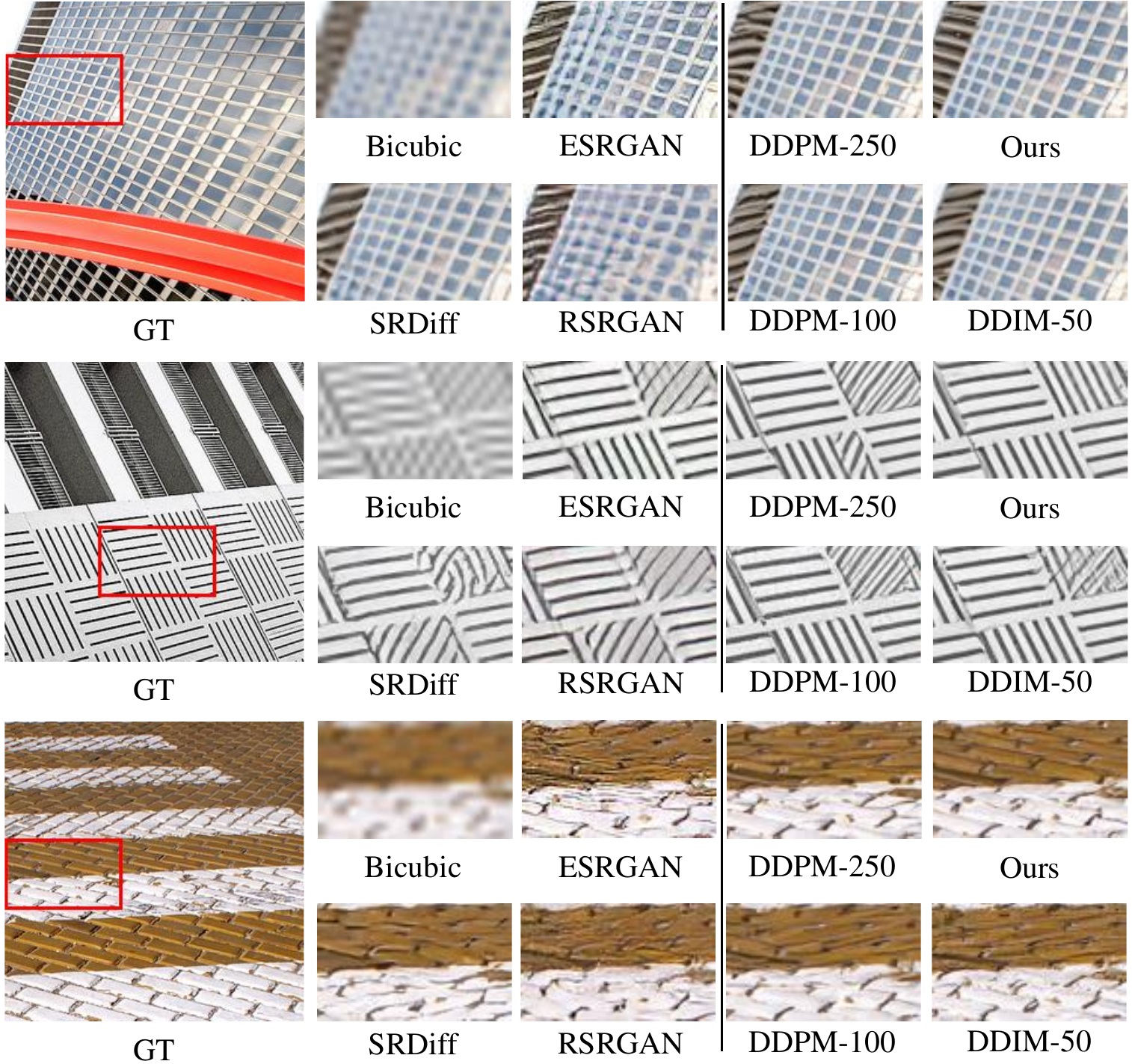}
    \caption{Qualitative comparisons of bicubic-SR results obtained by different methods. ``RSRGAN'' denotes RankSRGAN \citep{2019RankSRGAN}. All images on the right of the black line are sampled from the same vanilla diffusion-based SR model trained by us. \textbf{[Zoom in for best view]}}
    %\textcolor{red}{Red} numbers denote the best performances and \textcolor{blue}{blue} numbers denote the second best performances.}
    \label{fig: comparison}
    \vspace{-5mm}
\end{figure}

\subsection{Quantitative and Qualitative Results}
\label{subsec: comparisons}
The performances \myy{of bicubic-SR and real-SR} on testing datasets are shown in \myy{Tab.~\ref{tab: main bicubic qualitative results} and Tab.~\ref{tab: main real qualitative results} respectively}.
\myy{For bicubic-SR,} the performance of DDPM-1000 shows the capacity of the model, while the commonly-used sampling methods including DDPM-250, DDPM-100 trade off sample quality for faster sampling speed.
It can be seen that the performance of the proposed sampling method with \textit{diffusion ODE} solver of DDIM-100 outperforms all other sampling methods of the same diffusion-based SR model.
%including the previous upper-bound DDPM-1000 which is 20-times slower than our method. 
Remarkably, our method surpasses the previous upper-bound DDPM-1000, which is much slower.
\myy{For real-SR, our method surpasses the official sampling method of StableSR \citep{2023StableSR}, DDPM-200, with a faster sampling speed, unleashing the capability of StableSR better. 
%The performances of LPIPS of StableSR \citep{2023StableSR} on RealSR dataset are not competitive enough because the degradation of this dataset is not severe.
}
Such results demonstrate that we can steadily generate high-quality SR images from the pre-trained diffusion-based SR models by the proposed method. Visual comparisons of bicubic-SR images of different methods are shown in Fig.~\ref{fig: comparison}. More visual results can be found in the Sec.~\ref{suppsec: visual} of the appendix.

\subsection{Ablation Studies}
\label{subsec: ablation}

As we have discussed in the Sec.~\ref{subsec: approximating}, we use a reference set $\mathcal{R} = \{(\mathbf{z}_i, \mathbf{y}_i)\}_{i=1}^R$ and a set of randomly sampled $\mathbf{x}_T$ $\mathcal{K}$ to estimate the approximately optimal BC $\tilde{\mathbf{x}}_T$. The scales of the two sets will affect the quality of the estimated $\tilde{\mathbf{x}}_T$. The larger $\mathcal{R}$ and $\mathcal{K}$ are, the better estimation of $\tilde{\mathbf{x}}_T$ is. \myy{Thus, we perform ablation studies of the scale of the two sets on the task of bicubic-SR.}

\begin{figure}[htp]
    \centering
    \includegraphics[width=0.8\linewidth]{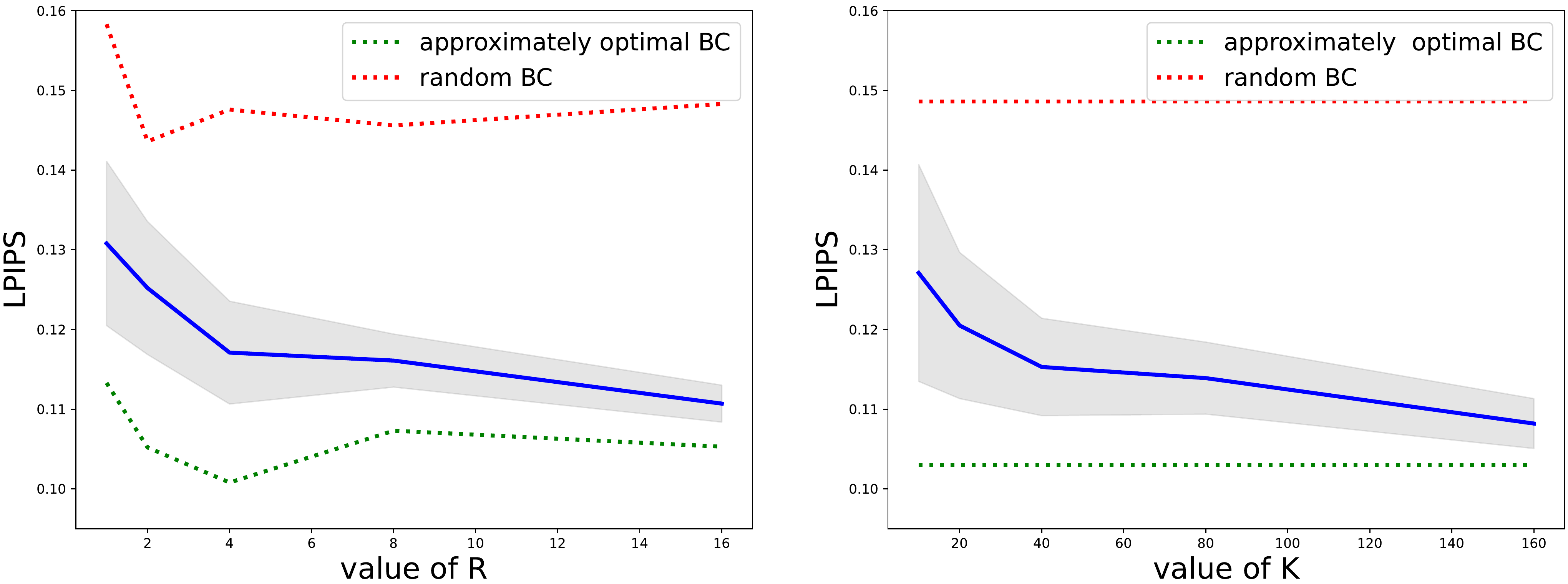}
    \caption{
    Ablation on values of $R$ and $K$. Shadows denote the standard deviation, the red dotted lines denote LPIPS of SR samples of the subset by DDIM-50 with randomly sampled $\mathbf{x}_T$, indicating the lower-bound of performance, and the green dotted lines denote LPIPS of SR results of the subset by DDIM-50 with $\tilde{\mathbf{x}}_T$, indicating the upper-bound of performance.
    }
    \label{fig: ablation}
    \vspace{-5mm}
\end{figure}

For the ablation on $\mathcal{R}$, we keep $K=200$. We build subsets $\mathcal{R}_i$ containing $i$ image pairs and set $i$ to 1, 2, 4, 8, 16. For each $i$, we build 8 $\mathcal{R}_i$ with different random image pairs. With the criterion of each $\mathcal{R}_i$, we choose the corresponding $\tilde{\mathbf{x}}_T$ and test them on a subset of DIV2k test set containing 100 patches with DDIM-50. The mean values and standard deviation values of LPIPS of the SR results with estimated $\tilde{\mathbf{x}}_T$ at each $i$ are shown in Fig.~\ref{fig: ablation}. 
% It can be seen that the performances become better and steadier as $R=i$ increases. 

For the ablation on $\mathcal{K}$, we keep $R=20$. We randomly sample $i$ $\mathbf{x}_T$ to build sets $\mathcal{K}_i$ and set i to 10, 20, 40, 80, 160. For each $i$, we build 8 $\mathcal{K}_i$ with different $\mathbf{x}_T$. We estimate $\tilde{\mathbf{x}}_T$ from each $\mathcal{K}_i$ and test them on the same subset of DIV2k test set used in the ablation studies on $\mathcal{R}$ with DDIM-50. The mean values and standard deviation values of LPIPS of the SR results with estimated $\tilde{\mathbf{x}}_T$ at each $i$ are \myy{also} shown in Fig.~\ref{fig: ablation}.

\myy{It can be seen that the performances become better and steadier as $R$ and $K$ increase.}
% It can be seen that the performances become better and steadier as $K=i$ increases. 

\section{Conclusion and Future Work}

In this work, we propose to steadily sample high-quality SR images from diffusion-based SR models by solving \textit{diffusion ODE}s with approximately optimal BCs $\tilde{\mathbf{x}}_T$.
We describe the process of finding these optimal boundary conditions.
Experiments show that the proposed sampling method outperforms commonly-used sampling methods for diffusion-based SR models. 
%Our method does not have any restrict on the architectures of the diffusion-based SR models and does not require any additional training, which means our method can be used to assist pre-trained diffusion-based SR models to sample better results.
%
Our method is not limited to specific architectures of diffusion-based SR models,  and does not require additional training. This flexibility allows our method to effectively enhance the sampling performance of pre-trained diffusion-based SR models without any constraints in a plug-and-play manner.

% In this work, we only discuss the SR tasks under the bicubic degradation. However, our analysis does not limit the formulation of tasks. In the future, we will manage to extend the proposed method to other low-level tasks including image colorization, low-light enhancement, blind image super-resolution, \textit{etc.} 
The calculated approximately optimal BC $\tilde{\mathbf{x}}_T$ has the same dimension as LR images $\mathbf{y}$, which can not be directly applied to LR images with other shapes. 
\myy{We will \wh{explore designing} algorithms which combine the $\tilde{\mathbf{x}}_T$ with resolution-arbitrary sampling methods \citep{2023DiffCollage} to achieve the goal of applying our method on LR images with different resolutions. 
Besides, we only discuss the $\tilde{\mathbf{x}}_T$ in the tasks of image super-resolution. In the future, we will explore further application in other low-level tasks, including image colorization, low-light enhancement, \textit{etc.}}
% We will manage to design algorithms to extend the application of $\tilde{\mathbf{x}}_T$ to LR images with other shapes.
%Besides, due to the limitation of existing diffusion-based SR methods, we have not outperformed \wh{SOTA} GAN-based SR methods yet. We will attempt to design better diffusion-based SR frameworks.
% ### I suggest not mentioning the last term. This drawback is very severe that might affect our score.

\section*{Acknowledgement}

This work was supported in part by the National Natural Science Foundation of China under Grant 62332010, and in part by the Key Laboratory of Science, Technology and Standard in Press Industry (Key Laboratory of Intelligent Press Media Technology).

\bibliography{iclr2024_conference}
\bibliographystyle{iclr2024_conference}

\newpage
\appendix
\setcounter{equation}{0}
\section*{Appendix}

\renewcommand\theequation{A.\arabic{equation}}
\section{Proof to Eqn.~\ref{eqn: re-sample}}
\label{suppsec: proof}
\textbf{Proposition.} \textit{The likelihood of $\mathbf{x}_T$ which is obtained by replacing the variable from $\mathbf{x}_0$ to $\mathbf{x}_T$ in the likelihood of $\mathbf{x}_0$ satisfies:}
\begin{equation}
\nonumber
    p_\theta'(\mathbf{x}_T|\mathbf{y})
    = p_\theta(h_\theta(\mathbf{x}_T, \mathbf{y})).
\end{equation}
\textbf{Proof.} First, we apply the law of total probability to all the choices of $\bar{\mathbf{y}} \in \mathcal{C}$, getting:
\begin{equation}
     p_\theta'(\mathbf{x}_T|\mathbf{y}) = \sum_{\bar{\mathbf{y}}\in\mathcal{C}} {p_\theta(\mathbf{x}_0|\mathbf{y})|_{\mathbf{x}_0 = h_\theta(\mathbf{x}_T, \bar{\mathbf{y}})}p(\bar{\mathbf{y}})},
\end{equation}
where $\mathcal{C}$ is the theoretically universal set of all LR images and $\bar{y}$ indicate all the LR images. If $\bar{\mathbf{y}}\neq\mathbf{y}$, $ p_\theta(h_\theta(\mathbf{x}_T, \bar{\mathbf{y}})|\mathbf{y})$ would indicate the probability of the generated image $h_\theta(\mathbf{x}_T, \bar{\mathbf{y}})$ being the corresponding SR image of another LR image $\mathbf{y}$, which is almost $0$. Thus, $\bar{\mathbf{y}}$ can only be equal to $\mathbf{y}$. Thus, we have:
\begin{equation}
    p_\theta(h_\theta(\mathbf{x}_T, \bar{\mathbf{y}})|\mathbf{y}) = \left\{
    \begin{aligned}
        &0, \bar{\mathbf{y}}\neq\mathbf{y}\\
        &p_\theta(h_\theta(\mathbf{x}_T, \mathbf{y})|\mathbf{y}), \bar{\mathbf{y}}=\mathbf{y}
    \end{aligned}
    \right.,
\end{equation}
furthermore,
\begin{equation}
    \sum_{\bar{\mathbf{y}}\in\mathcal{C}} {p_\theta(\mathbf{x}_0|\mathbf{y})|_{\mathbf{x}_0 = h_\theta(\mathbf{x}_T, \bar{\mathbf{y}})}p(\bar{\mathbf{y}})}
    =
    p_\theta(h_\theta(\mathbf{x}_T, \mathbf{y})|\mathbf{y})p(\mathbf{y})
    =
    p_\theta(h_\theta(\mathbf{x}_T, \mathbf{y}),\mathbf{y})
    .
\end{equation}
% When given $\bar{\mathbf{y}}=\mathbf{y}$, the condition of $p_\theta(h_\theta(\mathbf{x}_T, \mathbf{y})|\mathbf{y})$ can be removed because it is already given. Thus, we have:
Because $\mathbf{y}$ is the reverse function of $h_\theta(\mathbf{x}_T, \mathbf{y})$ which is deterministic, we have:
\begin{equation}
\label{eqn: full re-sample 1}
    \sum_{\bar{\mathbf{y}}\in\mathcal{C}} {p_\theta(\mathbf{x}_0|\mathbf{y})|_{\mathbf{x}_0 = h_\theta(\mathbf{x}_T, \bar{\mathbf{y}})}p(\bar{\mathbf{y}})}
    =
    p_\theta(h_\theta(\mathbf{x}_T, \mathbf{y})).
\end{equation}
Hence, the proof is completed.
\hfill $\square$

\setcounter{equation}{0}
\renewcommand\theequation{B.\arabic{equation}}

\section{Full Implementation Details of the SR3 baseline}
\label{suppsec: implementation}

The implementation details of the SR3 baseline trained by us contain two parts: details of the diffusion-based SR model $p_\theta(\mathbf{x}_0|\mathbf{y})$ and details of the noise-prediction network used by the diffusion model $\bm{\epsilon}_\theta(\mathbf{x}_t, \mathbf{y}, t)$. We provide them separately in this section, ensuring the reproducibility of our results.

\subsection{Implementation Details of the Diffusion Model}

We use the original diffusion model introduced by \cite{2020DDPM} which only predicts the noise in noisy state $\mathbf{x}_t$ without predicting the variances. Thus, the model can be simply trained through the mean square error (MSE) loss between the predicted noise and the real noise. The training loss is:
\begin{equation}
    \mathcal{L} = \mathbbm{E}_{t,\mathbf{x}_0,\bm{\epsilon}}||\bm{\epsilon} - \bm{\epsilon}_\theta(\mathbf{x}_t, \mathbf{y}, t)||^2,
\end{equation}
where $\mathbf{y}$ denotes LR images and $\bm{\epsilon}_\theta(\mathbf{x}_t, \mathbf{y}, t)$ is the noise-prediction network which is particularly depicted in the Sec.~\ref{subsec: implement unet}.
The noise schedule is the same as \cite{2020DDPM}, which sets $T$ to $1000$ and the forward process variances to constants increasing linearly from $\beta_1 = 10^{-4}$ to $\beta_T = 0.02$. During the reverse process of DDPM, we set the variance $\sigma_t$ to $\frac{1-\bar{\alpha}_{t-1}}{1-\bar{\alpha}_t}\beta_t$ which performs much better than $\sigma_t = \beta_t$ in resampled few-step sampling, following \cite{2021ImprovedDDPM}. Following \cite{2021ADM}, we use a resampled schedule for a few-step sampling. For DDPM-250, we use the schedule of $90,60,60,20,20$, which is the same as the best schedule for image generation tasks found by \cite{2021ADM}. For DDPM-100, we use the schedule of $45,20,15,10,10$, which is not exhaustively swept.

\subsection{Implementation Details of the Noise-Prediction Network}
\label{subsec: implement unet}

Following most of the current diffusion models \citep{2020DDPM, 2021SR3, 2021SRDiff, 2023ResDiff, 2022MMDiffusion, 2023UMMDiffusion, 2022DALLE2, 2022Imagen, 2022StableDiffusion} used in several aspects, we use UNet as the backbone of our noise-prediction network. Following SR3 \citep{2021SR3}, the LR images $\mathbf{y}$ are first upsampled by the bicubic kernel to the same size to noise states $\mathbf{x}_t$ and then simply concatenated to noise states $\mathbf{x}_t$ along the channel dimension. The bicubic kernel we used both in downsampling and upsampling is introduced by torchvision \citep{2023PyTorch} with anti-alias. The architecture of our UNet is similar to the upsampler built by \cite{2021ADM} with a small number of parameters. The detailed architecture is shown in Tab.~\ref{tab: Unet details}. We first train the model for 2M iterations with a batch size of 16, then train the model for another 1M iterations with a batch size of 64, ensuring the convergence of our model. We use Adam optimizer \citep{2014Adam} during the whole training process and use mixed precision to accelerate training. The total training cost is about 2000 Tesla V100 GPU$\cdot$hours.
\begin{table}[htbp]
    \caption{Detailed architecture of our UNet used for the diffusion-based SR model.}
    \centering
    \begin{tabular}{cc}
        \toprule
         & UNet $64\rightarrow256$ \\
        \midrule 
        Model size & 36M 
        \\ 
        Channels & 92
        \\
        Depth & 2
        \\
        Channels multiple & 1,1,2,2,3
        \\
        Heads & 4
        \\
        Attention resolution & 32,16
        \\
        BigGAN up/downsample & \checkmark
        \\
        Dropout & 0.0
        \\
        Batch size & $16\rightarrow64$
        \\
        Iterations & 2M + 1M
        \\
        Learning rate & $1e-4$
        \\
        \bottomrule
    \end{tabular}
    \label{tab: Unet details}
\end{table}

\section{The Sources of the Compared Methods}
\label{suppsec: code source}
The sources of the compared methods including SRDiff \citep{2021SRDiff}, ESRGAN \citep{2018ESRGAN}, RankSRGAN \citep{2019RankSRGAN}, RealSR \citep{2020RealSR}, BSRGAN \citep{2021BSRGAN}, Real-ESRGAN \cite{2021RealESRGAN}, DASR \citep{2022DASR}, and KDSR-GAN \cite{2023KDSR} are shown in Tab.~\ref{tab: sources}. It is noted that the model of RealSR \citep{2020RealSR} we employ is ``DF2K-JPEG''.

\begin{table}[ht]
    \caption{The sources of the compared methods.}
    \centering
    \begin{tabular}{c|c|c}
    \toprule
         Degradation & Method & URL
         \\
         \midrule
         \multirow{3}{*}{Bicubic} & SRDiff & \url{https://github.com/LeiaLi/SRDiff}
         \\
         & ESRGAN & \url{https://github.com/xinntao/ESRGAN}
         \\
         & RankSRGAN & \url{https://github.com/XPixelGroup/RankSRGAN}
         \\
         \midrule
         \multirow{5}{*}{Real} & RealSR & \url{https://github.com/jixiaozhong/RealSR}
         \\
         & BSRGAN & \url{https://github.com/cszn/BSRGAN}
         \\
         & Real-ESRGAN & \url{https://github.com/xinntao/Real-ESRGAN}
         \\
         & DASR & \url{https://github.com/csjliang/DASR}
         \\
         & KDSR-GAN & \url{https://github.com/Zj-BinXia/KDSR}
         \\
    \bottomrule
    \end{tabular}
    \label{tab: sources}
\end{table}

\section{Boosting Mid-Training Models}

In the main paper, we analyze all the characteristics of $\mathbf{x}^*_T$ and propose method of approximating $\tilde{\mathbf{x}}_T$ assuming the diffusion-based SR model has been well-trained (\ie $p_\theta(\mathbf{x}_0|\mathbf{y})$ is a close fit to $q(\mathbf{x}_0|\mathbf{y})$). However, we find that the proposed method can also boost mid-training models. We use the baseline SR3 model of bicubic-SR. The model is trained for only 500k iterations with a batch size of 16, costing 200 Tesla V100 GPU$\cdot$hours. The performances are shown in Tab.~\ref{tab: mid-training performance}. 
\begin{table}[htp]
    \caption{Performances of the mid-training model with only 500k training iterations. \textcolor{red}{Red} numbers denote the best performances among the mid-training model and \textcolor{blue}{blue} numbers denote the second best performances among the mid-training model.}
    \centering
    \begin{tabular}{c|c|cc|cc|cc}
        \toprule
        \multirow{2}{*}{Model} & \multirow{2}{*}{\makecell{Sampling\\method}} & \multicolumn{2}{c|}{DIV2k-test} & \multicolumn{2}{c|}{Urban100} & \multicolumn{2}{c}{BSD100}
        \\
         & & LPIPS~$\downarrow$ & PSNR~$\uparrow$ & LPIPS~$\downarrow$ & PSNR~$\uparrow$ & LPIPS~$\downarrow$ & PSNR~$\uparrow$
        \\
        \midrule
        \multirow{5}{*}{\makecell{well-\\trained}} & DDPM-1000 & 0.1075 & 28.75 & 0.1165 & 24.33 & 0.1555 & 23.86
        \\
         & DDPM-250 & 0.1142 & 28.95 & 0.1181 & 24.41 & 0.1621 & 24.00
        \\
         & DDPM-100 & 0.1257 & 29.16 & 0.1232 & 24.51 & 0.1703 & 24.15
        \\
         & DDIM-50 & 0.1483 & 28.55 & 0.1333 & 24.16 & 0.1823 & 23.75
        \\
         & DDIM-50 + $\tilde{\mathbf{x}}_T$ &  0.1053 & 28.65 & 0.1164 & 24.26 & 0.1552 & 23.99
        \\
        \midrule
        \midrule
        \multirow{5}{*}{\makecell{mid-\\training}} & DDPM-1000 & 0.2403 & 18.57 & 0.1663 & \textcolor{blue}{19.34} & 0.2269 & 18.77
        \\
         & DDPM-250 & 0.2361 & 18.65 & 0.1734 & 19.05 & \textcolor{blue}{0.2249} & 18.81
        \\
         & DDPM-100 & \textcolor{blue}{0.2315} & \textcolor{blue}{18.71} & \textcolor{blue}{0.1640} & 19.30 & \textcolor{red}{0.2140} & \textcolor{blue}{18.98}
        \\
         & DDIM-50 & 0.4536 & 17.10 & 0.3098 & 17.62 & 0.4040 & 17.84
         \\
         & DDIM-50 + $\tilde{\mathbf{x}}_T$ & \textcolor{red}{0.2209} & \textcolor{red}{19.15} & \textcolor{red}{0.1618} & \textcolor{red}{19.97} & 0.2514 & \textcolor{red}{20.49}
        \\
        \bottomrule
    \end{tabular}
    \label{tab: mid-training performance}
\end{table}
We suspect that the reason for the boosting of the mid-training model is although the mid-training model is not a close fit to $q(\mathbf{x}_0|\mathbf{y})$ yet, it has learned the extreme points of $q(\mathbf{x}_0|\mathbf{y})$. Thus, the assumptions corresponding to extreme points approximately hold (\ie Eqn.~\ref{eqn: optimal x_0}, Eqn.~\ref{eqn: max}). So, we still can extract a $\tilde{\mathbf{x}}_T$ based on Eqn.~\ref{eqn: target} and use it as an approximately optimal BC to other LR images $\mathbf{y}$, getting better performances. We observe that DDIM-50 performs much worse than other sampling methods when applied to the mid-training diffusion-based SR model. Such phenomenon is in conflict with the conclusion of applying these sampling methods in diffusion-based image generation models \citep{2021ImprovedDDPM}. However, our method can still boost the DDIM-50 (\ie the \textit{diffusion ODE} solver used in the paper) with the approximately optimal BC $\tilde{\mathbf{x}}_T$, reaching comparable performances with DDPM-based sampling methods.

\begin{table}[htbp]
    \caption{Pearson's coefficients between 10 LPIPS sequences of 100 bicubic-SR images for each LR image generated by the baseline SR3 model.}
    \centering
    \begin{tabular}{c|cccccccccc}
        \toprule
         & LR-1 & LR-2 & LR-3 & LR-4 & LR-5 & LR-6 & LR-7 & LR-8 & LR-9 & LR-10
         \\
         \midrule
         LR-1 & 1.000 & 0.754 & 0.840 & 0.798 & 0.811 & 0.751 & 0.902 & 0.837 & 0.877 & 0.765
         \\
         LR-2 & 0.754 & 1.000 & 0.811 & 0.789 & 0.832 & 0.702 & 0.831 & 0.775 & 0.812 & 0.717
         \\
         LR-3 & 0.840 & 0.811 & 1.000 & 0.732 & 0.799 & 0.654 & 0.841 & 0.799 & 0.836 & 0.745
         \\
         LR-4 & 0.798 & 0.789 & 0.732 & 1.000 & 0.756 & 0.699 & 0.855 & 0.801 & 0.793 & 0.732
         \\
         LR-5 & 0.811 & 0.832 & 0.799 & 0.756 & 1.000 & 0.632 & 0.811 & 0.792 & 0.856 & 0.789
         \\
         LR-6 & 0.751 & 0.702 & 0.654 & 0.699 & 0.632 & 1.000 & 0.721 & 0.734 & 0.611 & 0.704
         \\
         LR-7 & 0.902 & 0.831 & 0.841 & 0.855 & 0.811 & 0.721 & 1.000 & 0.754 & 0.787 & 0.725
         \\
         LR-8 & 0.837 & 0.775 & 0.799 & 0.801 & 0.792 & 0.734 & 0.754 & 1.000 & 0.813 & 0.786
         \\
         LR-9 & 0.877 & 0.812 & 0.836 & 0.793 & 0.856 & 0.611 & 0.787 & 0.813 & 1.000 & 0.801
         \\
         LR-10 & 0.765 & 0.717 & 0.745 & 0.732 & 0.789 & 0.704 & 0.725 & 0.786 & 0.801 & 1.000
         \\
         \bottomrule
    \end{tabular}
    \label{tab: pearson matrix}
\end{table}

\section{Validation on the Independence of \texorpdfstring{$p_\theta(h_\theta(\mathbf{x}_T, \mathbf{y}))$}{} to \texorpdfstring{$\mathbf{y}$}{}}
\label{suppsec: validation}

\begin{figure}
    \centering
    \includegraphics[width=0.75\linewidth]{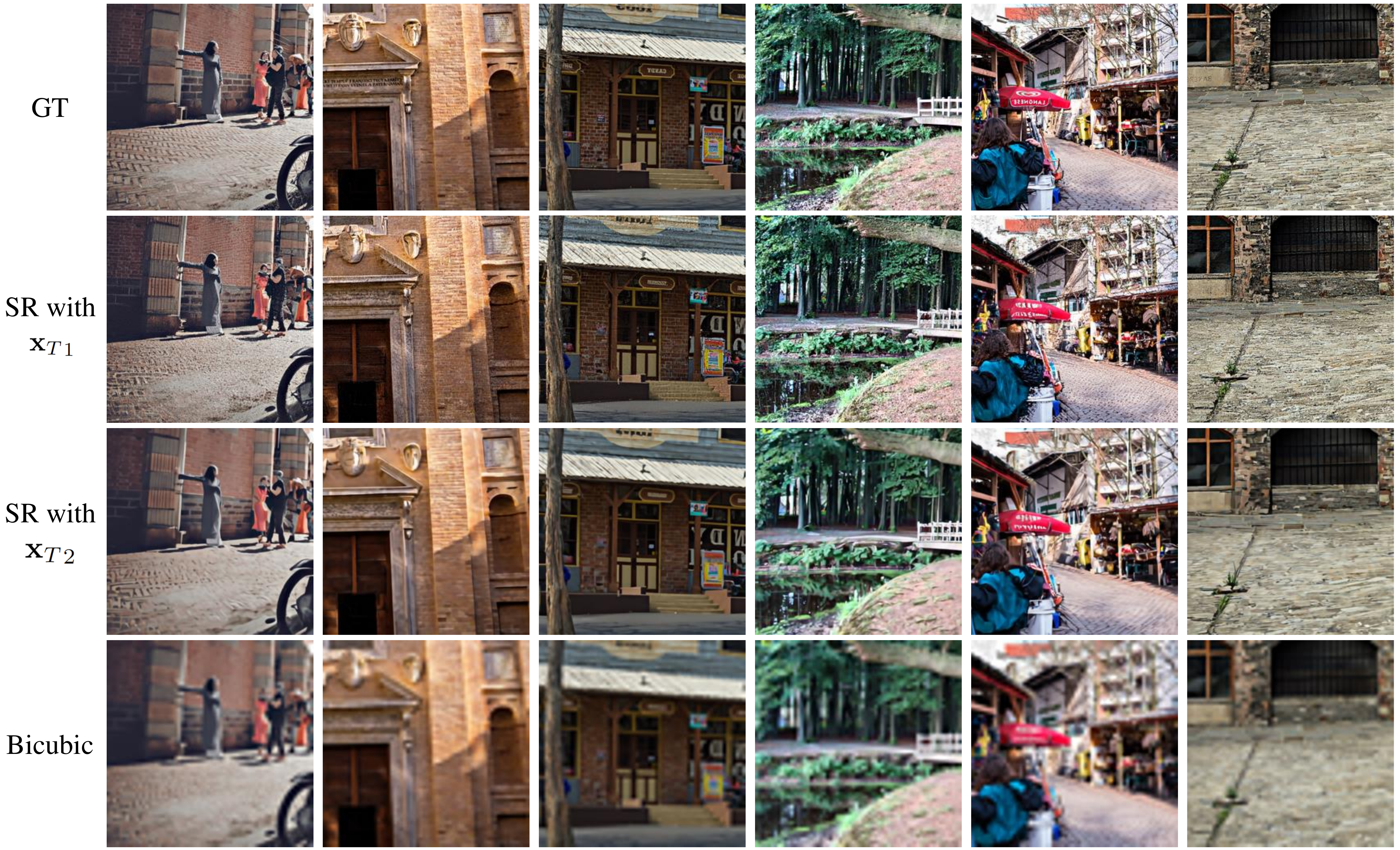}
    \caption{SR results with shared $\mathbf{x}_T$.
    Results with ${\mathbf{x}_T}_1$ all have excessive artifacts and results with ${\mathbf{x}_T}_2$ are all over-smooth. 
    Results with shared $\mathbf{x}_T$ share visual features. \textbf{[Zoom in for best view]}}
    \label{fig: shared x_T}
\end{figure}

As we have stated in the Sec.~\ref{subsec: analyzing}, $p_\theta(h_\theta(\mathbf{x}_T, \mathbf{y}))$ is not related to the specific LR images $\mathbf{y}$. 
In this section, we design an experiment to show the related evidence.
As we have mentioned in the Sec.~\ref{subsec: approximating}, we assume distance measurement function $M(h_\theta(\mathbf{x}_T, \mathbf{y}), \mathbf{z})$ has the same shape as $q(h_\theta(\mathbf{x}_T, \mathbf{y}))$ and we use $q(h_\theta(\mathbf{x}_T, \mathbf{y}))$ to approximate $p_\theta(h_\theta(\mathbf{x}_T, \mathbf{y}))$. So, given different LR images $\mathbf{y}_i$, if $p_\theta(h_\theta(\mathbf{x}_T, \mathbf{y}_i))$ are independent, the functions $M(h_\theta(\mathbf{x}_T, \mathbf{y}_i), \mathbf{z}_i)$ of $\mathbf{x}_T$ should have the same shape.
Thus, we validate the shapes of $M(h_\theta(\mathbf{x}_T, \mathbf{y}_i), \mathbf{z}_i)$ of different $\mathbf{y}_i$. We randomly sample 10 bicubic-LR image pairs and 100 $\mathbf{x}_T$, then generate 100 SR images by the baseline SR3 model of each LR image and calculate their LPIPS, getting 10 LPIPS sequences. 
To evaluate the shapes of the 10 LPIPS sequences, we calculate the Pearson correlation coefficients of every two sequences and form a matrix shown in Tab.~\ref{tab: pearson matrix}. It can be seen that the coefficients are all high, indicating the strong correlation between different LPIPS sequences. 
To visualize the correlation between SR results of different LR images $\mathbf{y}_i$, we further exhibit several SR images sharing the same $\mathbf{x}_T$ in Fig.~\ref{fig: shared x_T}.
It can be seen that SR images of different LR images with the same $\mathbf{x}_T$ have similar visual features. SR results with ${\mathbf{x}_T}_1$ seem over-sharp and contain excessive artifacts while SR results with ${\mathbf{x}_T}_2$ seem over-smooth. All of them are reasonable but not satisfying enough, indicating the necessity of finding an approximately optimal BC $\tilde{\mathbf{x}}_T$.

It should be noticed that this experiment only validates that the consistency of shapes of $M(h_\theta(\mathbf{x}_T, \mathbf{y}_i), \mathbf{z}_i)$, which is an derivation of the independence of $p_\theta(h_\theta(\mathbf{x}_T, \mathbf{y}))$ to $\mathbf{y}$, instead of the independence itself.

\section{More Visual Results}
\label{suppsec: visual}

\begin{figure}[htbp]
    \centering
    \includegraphics[width=0.75\linewidth]{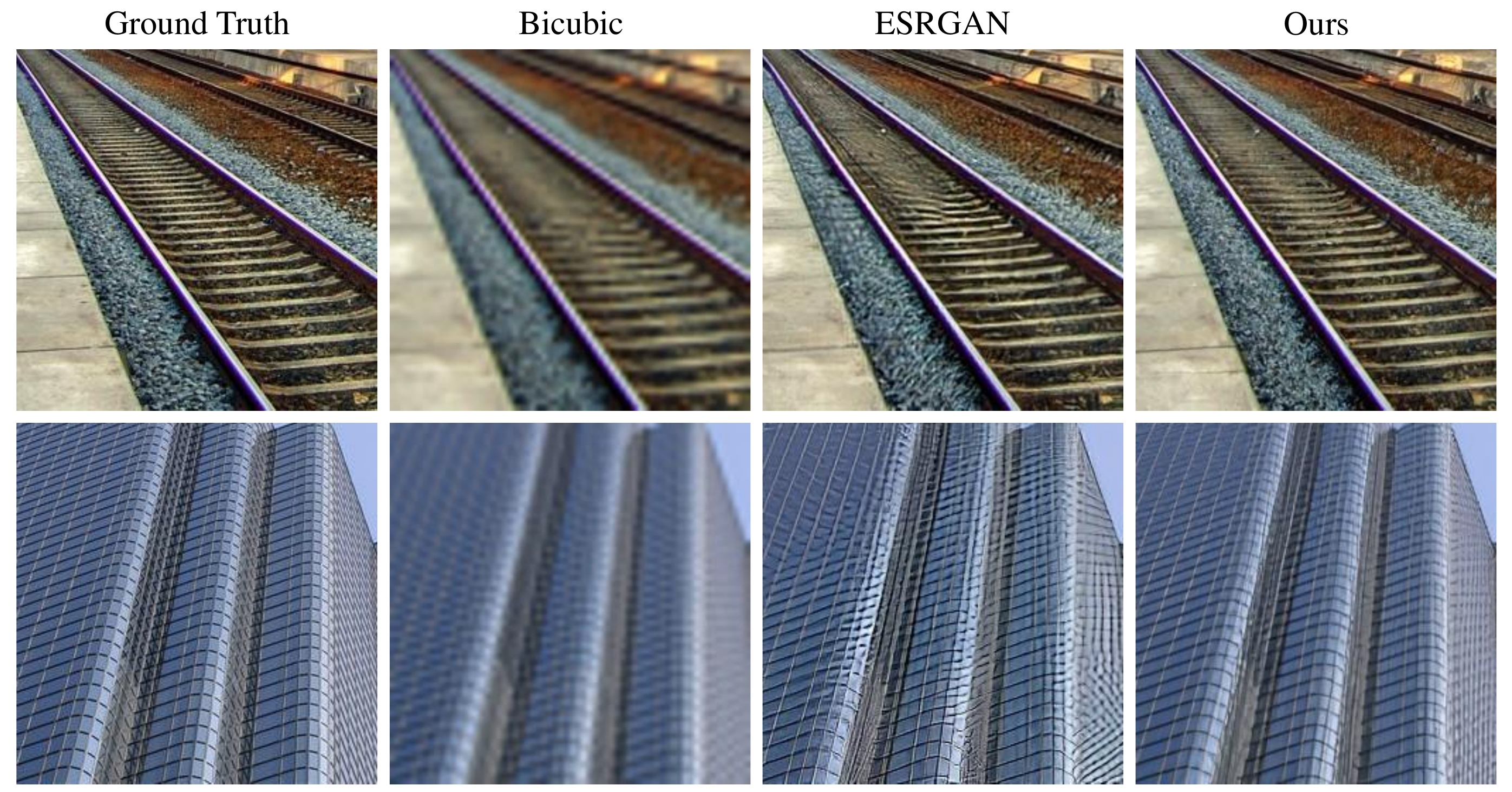}
    \caption{Further visual comparisons. \textbf{[Zoom in for best view]}}
    \label{fig: supp result1}
\end{figure}

\begin{figure}[htbp]
    \centering
    \includegraphics[width=\linewidth]{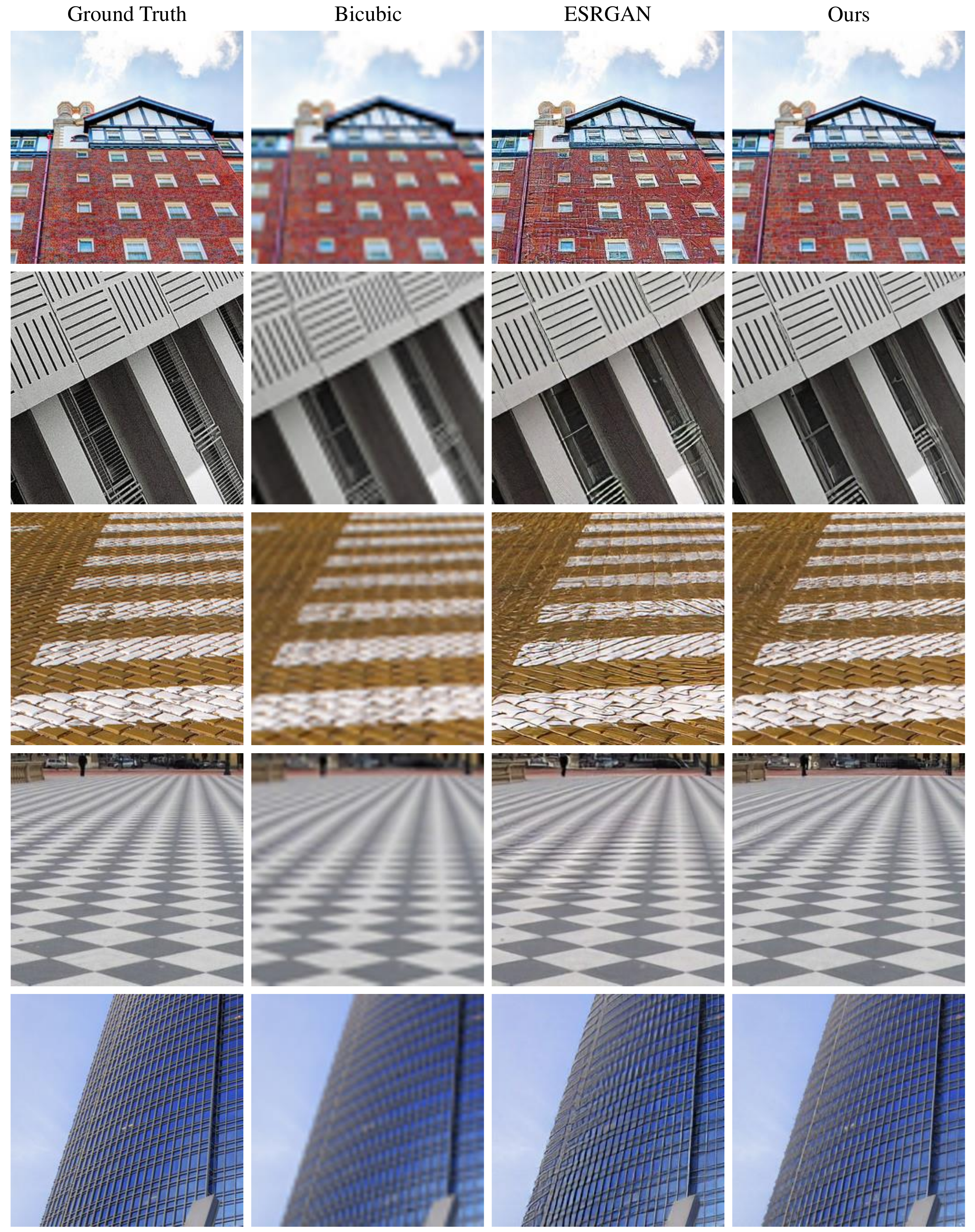}
    \caption{Further visual comparisons. \textbf{[Zoom in for best view]}}
    \label{fig: supp result2}
\end{figure}

\begin{figure}[htbp]
    \centering
    \includegraphics[width=\linewidth]{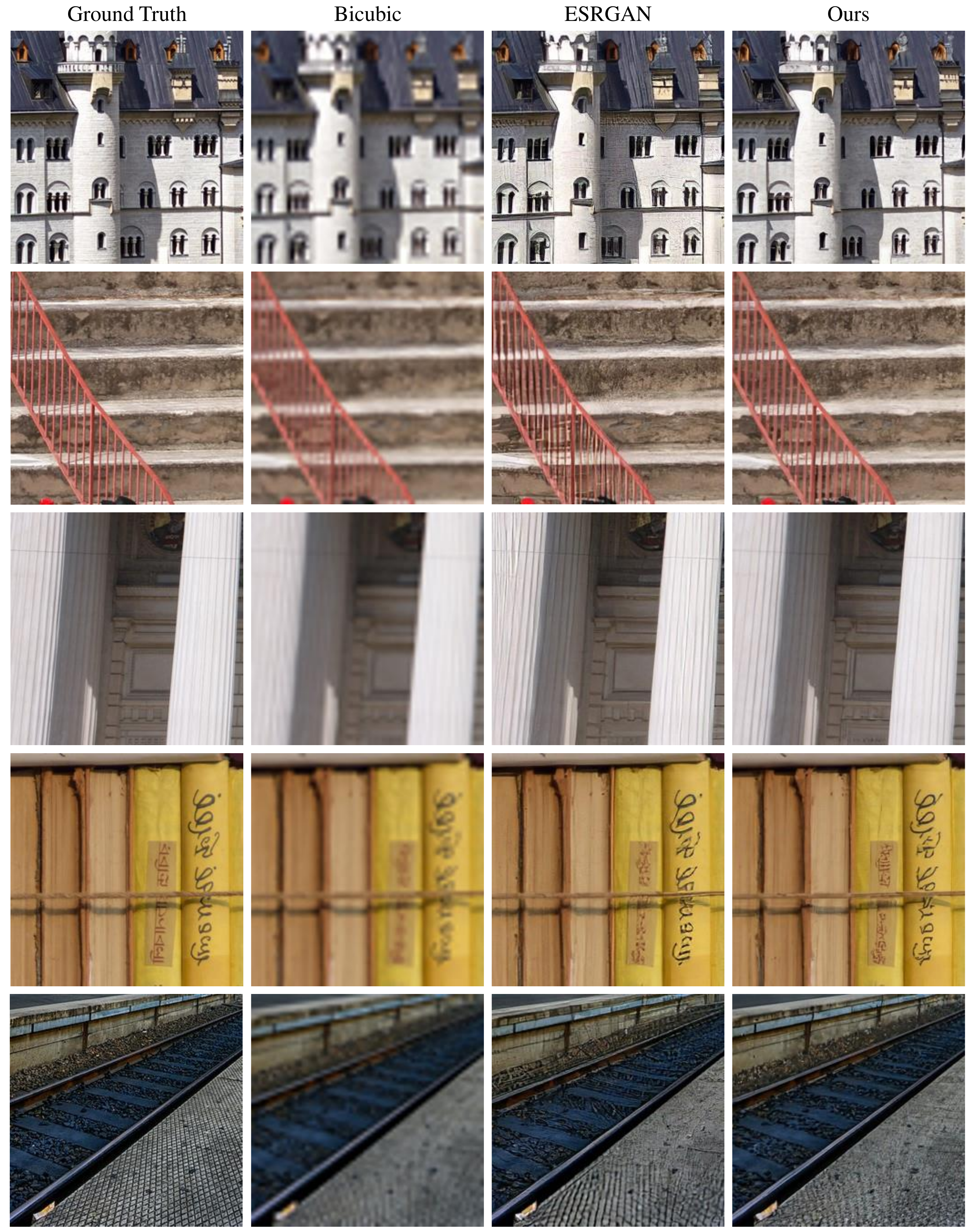}
    \caption{Further visual comparisons. \textbf{[Zoom in for best view]}}
    \label{fig: supp result3}
\end{figure}

In this section, we show more visual results of bicubic-SR compared with ESRGAN \citep{2018ESRGAN} (which is the representative of GAN-based methods) in Fig.~\ref{fig: supp result1}, Fig.~\ref{fig: supp result2} and Fig.~\ref{fig: supp result3}, demonstrating the superiority of our method in perceptual quality.

\end{document}